# Imaging the Two Gaps of the High-$T_C$ Superconductor Pb-$Bi_2Sr_2CuO_{6+x}$

M. C. Boyer[1], W. D. Wise[1], Kamalesh Chatterjee[1], Ming Yi[1], Takeshi Kondo[2,1,*], T. Takeuchi[2,3], H. Ikuta[2], E. W. Hudson[1]

[1] Department of Physics, Massachusetts Institute of Technology, Cambridge, MA 02139, USA.
[2] Department of Crystalline Materials Science, Nagoya University, Nagoya 464-8603, Japan.
[3] EcoTopia Science Institute, Nagoya University, Nagoya 464-8603, Japan.
* Present address: Ames Laboratory and Department of Physics and Astronomy, Iowa State University, Ames, IA 50011, USA.

**The nature of the pseudogap state, observed above the superconducting transition temperature $T_C$ in many high temperature superconductors, is the center of much debate. Recently, this discussion has focused on the number of energy gaps in these materials. Some experiments indicate a single energy gap, implying that the pseudogap is a precursor state. Others indicate two, suggesting that it is a competing or coexisting phase. Here we report on temperature dependent scanning tunneling spectroscopy of Pb-$Bi_2Sr_2CuO_{6+x}$. We have found a new, narrow, homogeneous gap that vanishes near $T_C$, superimposed on the typically observed, inhomogeneous, broad gap, which is only weakly temperature dependent. These results not only support the two gap picture, but also explain previously troubling differences between scanning tunneling microscopy and other experimental measurements.**

Scanning tunneling microscopy (STM), with its ability to measure atomically resolved differential conductance spectra proportional to the local density of states of a material, has made important contributions to our current understanding of high temperature superconductivity (HTS)[1]. STM has also produced several controversial results, two of which we focus on here. First, STM spectra clearly show a gap in the density of states which smoothly evolves from the superconducting to the pseudogap state, at least in the most commonly studied Bi-based cuprates – $Bi_2Sr_2CaCu_2O_{8+x}$ (Bi-2212)[2] and $Bi_2Sr_2CuO_{6+x}$ (Bi-2201)[3]. In the overdoped $(Bi_{1-y}Pb_y)_2Sr_2CuO_{6+x}$ ($T_C$ = 15 K) samples used in this study[4], we observe this same smooth evolution (Fig. 1). The lack of any obvious change at $T_C$ in the STM-measured gap has been interpreted as an important piece of evidence for the single gap picture of HTS, and hence for theories that the pseudogap shares the pairing of the superconducting state, but lacks its phase coherence[5-11].

A second controversial result is nanoscale inhomogeneity, in which gap magnitudes are observed to vary wildly on nanometer length scales[12-16]. In STM, a spectral survey, in which the local density of states is measured at a dense array of locations, allows the mapping of spatial variations of spectral features. Such surveys have led to the direct visualization of atomic scale effects, such as single atom impurities[17,18] and oxygen dopant atoms[16]. To study inhomogeneity we extract from the survey a gap map $\Delta(\vec{r})$, where $\Delta$ is half the distance between the two edges of the gap (Fig. 2). Similar to previous measurements[12-16] we find wide gap magnitude variations on a nanometer length scale, which have been interpreted as due to variations in local superconducting pairing strength[16,19].

As mentioned, both of these interpretations of STM results are controversial. In the case of the single gap hypothesis, Deutscher has shown that the gap measured by STM and angle-resolved photoemission spectroscopy (ARPES) is different from the gap measured by Andreev reflection, penetration depth and Raman spectroscopy, and thus that there are two distinct energy scales in the system[20]. Even Nernst effect studies which, like STM, find a smooth thermal evolution (here indicating the presence of vortex fluctuations above $T_C$) do find an onset temperature which scales with $T_C$[21], rather than simply decreasing linearly with doping as the tunneling measured gap does[22]. In the case of inhomogeneity, Loram *et al*, for example, have noted that large superconducting gap variations on short length scales are inconsistent with NMR and heat capacity data[23]. Even within the STM community its interpretation is debated, particularly by Fischer *et al.* who strongly argue that inhomogeneity is due merely to stoichiometric disorder[1] and likely irrelevant to superconductivity.

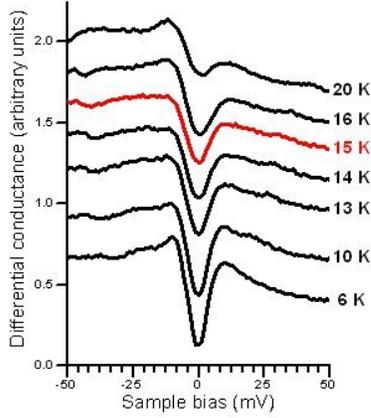

**Figure 1.** Temperature dependence of spatially averaged differential conductance spectra on one of several indistinguishable $(Bi_{1-y}Pb_y)_2Sr_2CuO_{6+x}$ (y = 0.38) samples used in this study, showing a smooth transition from the superconducting to pseudogap state at $T_C$ = 15 K (red). Spectra were measured using a standard lock-in technique and are shown with even vertical offsets to enhance visibility. All spectra in this paper were recorded with the same settings: $V_{Sample}$ = -100 mV, I = 100 pA, $V_{mod, rms}$ = 1.6 mV, and will be similarly offset.

In order to further investigate these and other phenomena, we have constructed an ultrahigh vacuum STM with the novel ability to measure spectral surveys of a selected, atomically resolved region of a sample over a wide range of temperatures. This ability is crucial in tracking the behavior of inhomogeneous spectra, as without it only average spectral temperature dependence may be studied. Using this technique, we are able to make gap maps of the same atomically resolved region as a function of temperature (Fig. 3). In a perhaps unsurprising extension of the smooth evolution from superconducting to pseudogap state shown in Fig. 1, we find that the gap maps are roughly independent of temperature, even when warmed through $T_C$ = 15 K. That is, in the pictured region, over fifteen thousand widely varying spectra evolve smoothly with temperature, apparently disregarding the superconducting transition at $T_C$, and thus preserving the initial gap width inhomogeneity.

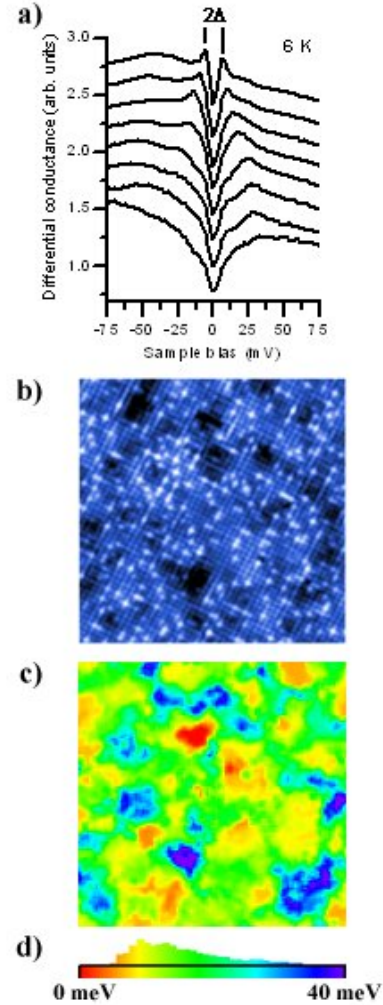

**Figure 2.** (a) A set of spectra associated with gap values ranging from Δ = 7 meV to more than 50 meV, where Δ is defined as half the distance between the edges of the gap. (b) 180 Å square topography showing Bi and Bi-replaced Pb atoms (the brighter atoms). We see no correlation between spectroscopy and the presence or absence of Pb atoms. (c) A gap map $\Delta(\vec{r})$ of the same region shown in (b), demonstrating large variations (Δ = 16 ± 8 meV), a histogram of which is shown in (d). Red (Δ = 0 meV) means that no gap was detected, and typically indicates the presence of an impurity state.

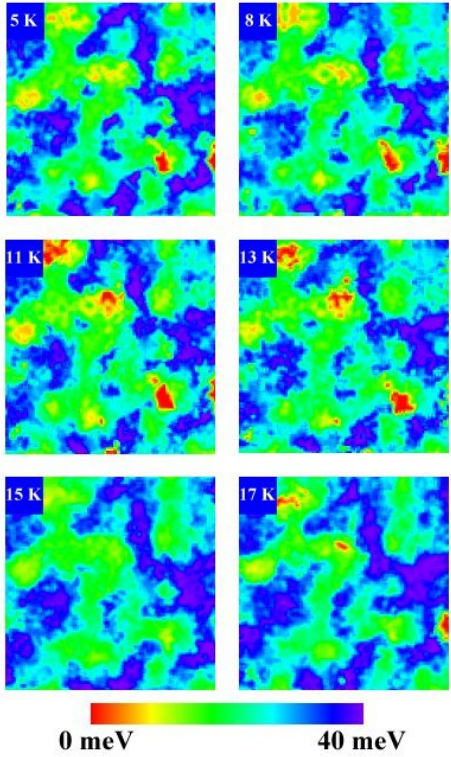

**Figure 3.** 175 Å gap maps derived from spectral surveys taken in the same location at T = 5 K, 8 K, 11 K, 13 K, 15 K and 17 K respectively. All maps share the color bar. Temperature independence of the inhomogeneity suggests that it is irrelevant to superconductivity, which vanishes at $T_C$ = 15 K in this sample.

The striking temperature independence of inhomogeneity, unaffected by the onset of superconductivity at $T_C$, led us to search for a signature of superconductivity by removing the effective background of the high temperature spectra from the low temperature ones. We thus calculate a normalized differential conductance $G_N$ as a function of energy $E$, position $\vec{r}$, and temperature $T$ by the common technique of dividing out the background, in this case a spectrum from the same position at a higher "normalization temperature" $T_N > T_C$:

$$G_N(E, \vec{r}, T) = G(E, \vec{r}, T) / G(E, \vec{r}, T_N)$$

We use division here for two reasons. First, it is often the correct normalization scheme, as, for example, in conventional (BCS) superconductors where the onset of superconductivity creates a superconducting density of states $N_S$ by opening a gap which multiplies the normal state density of states $N_0$:

$$N_S \approx N_0 \frac{|E|}{\sqrt{E^2 - \Delta^2}}$$

Second, other normalization schemes, such as subtraction, are difficult given that STM measured differential conductance is only proportional to the density of states, where the constant of proportionality is unknown as well as temperature and position dependent.

We show the result of this normalization on the set of spectra from Fig. 2a in Fig. 4a. After division, the temperature independent, inhomogeneous gap is removed and we find that a small gap remains. Applying the same normalization technique to the entire survey behind the gap map of fig. 2c demonstrates that this small gap is present throughout the sample, and is significantly more homogeneous than the larger gap we normalized away (Fig. 4b). The mean and standard deviation of measured gap magnitudes (Fig. 2c and Fig. 4b) drop from $\Delta_{large}$ = 16 ± 8 meV to $\Delta_{small}$ = 6.7 ± 1.6 meV.

Not only is this newly revealed small gap homogeneous, it must also have a different temperature dependence than the large gap that we normalized away. In order to clarify this, we plot (Fig. 5a) the average of normalized ($T_N$ = 17 K) spectral surveys at several temperatures below $T_C$ = 15 K. In contrast to the apparent temperature independence of the unnormalized spectra, we find that the normalized spectra are strongly temperature dependent, with the small gap vanishing near $T_C$.

One might protest that by choosing a normalization temperature $T_N$ close to $T_C$ we enforce this disappearance of the small gap. After all, $G_N(T = T_N)$ must be a straight line. In fact, the above results are relatively insensitive to our choice of $T_N$. In Fig. 5b we show that low temperature (T < $T_C$) spectra normalize to the same small gap regardless of $T_N$, while high temperature (T > $T_C$) spectra do not. That is, the small gap is present below $T_C$ but not above it. We choose to work with $T_N$ close to $T_C$ because the larger gap is not completely temperature independent and because at higher temperatures thermal broadening begins to obscure the picture.

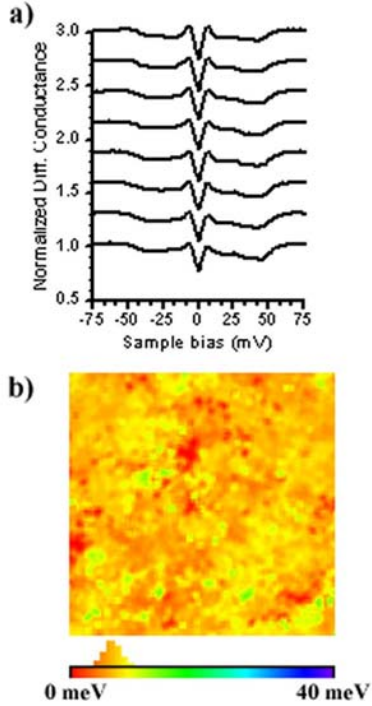

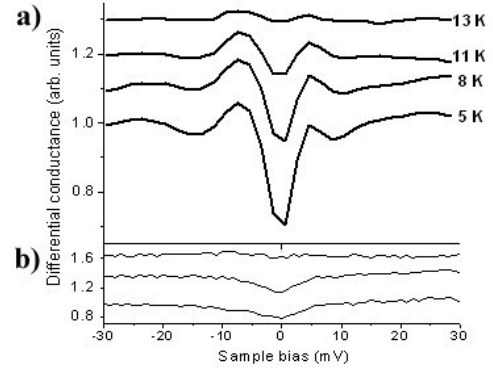

**Figure 4.** (a) A set of normalized spectra $G_N = G(6\,K)/G(16\,K)$ taken at the same locations as those shown in Fig. 2a. Note that all high energy variation is removed, leaving a small, consistent gap. (b) Gap map of normalized $T = 6$ K spectra from the same region as shown in Fig. 2, showing significantly increased homogeneity ($\Delta_N = 6.7 \pm 1.6$ meV). The small remaining $\Delta_N$ variations show no correlation with $\Delta_{6\,K}$ or $\Delta_{16\,K}$.

**Figure 5.** (a) Temperature dependence of spatially averaged, normalized spectra at $T = 5$ K, 8K, 11 K and 13 K. The gap vanishes near $T_C = 15$ K. Without an accurate fitting function it is difficult to determine whether the gap closes ($\Delta(T\rightarrow T_C)\rightarrow 0$) or fills ($\Delta(T) = \Delta_0$, $G_N(|E| < \Delta_0$, $T\rightarrow T_C) \rightarrow 1$), as thermal broadening always leads to the impression of a filling gap (where $\Delta$ appears to remain constant or even broadens) even when the gap is closing. (b) A $T = 13$ K spectrum normalized at $T_N = 16$ K (bottom) and $T_N = 19$ K (middle) shows insensitivity of our results to the choice of $T_N$, as both curves show a small gap, in contrast to spectra taken above $T_C$, which normalize to an ungapped spectrum (top).

A natural interpretation of these results is that the gap revealed by normalization, which is homogeneous and vanishes near $T_C$, is the true superconducting gap and coexists with the large inhomogeneous gap. The large gap is characteristic of a state (likely the pseudogap state) that develops at some temperature above $T_C$ and exists unperturbed down to $T = 0$ K (or at least below our measurement temperatures). It is worth noting that even though the samples we are working with are well overdoped, in Bi-2201 the pseudogap phase, which is typically associated with underdoped materials, has been observed to exist in this part of the phase diagram[3, 24, 25]. This coexistence is similar to that observed in the conventional superconductor $NbSe_2$, where a superconducting gap that appears at $T_C = 7.4$ K is superimposed on a bowl shaped charge density wave gap that opens at $T_{CDW} = 35$ K, and the resulting spectra are the multiplicative product of the two effects. This interpretation may explain the low energy homogeneity observed in even very inhomogeneous samples[15, 26, 27], as low energy behavior is dominated by the homogeneous superconducting gap. It also explains why probes of low energy excitations, such as heat capacity measurements[23], as well as STM measurements of quasi-particle interference patterns[26, 28], are impervious to high energy inhomogeneity.

Furthermore, our results are consistent with recent ARPES measurements demonstrating the existence of two distinct gaps in both deeply underdoped Bi-2212[29] and in optimally doped Bi-2201[30]. Both of these studies found that in regions of momentum space near the antinode, spectra are characterized by a large gap that is insensitive to warming through $T_C$, while near the node, spectra have a narrow gap which closes at $T_C$. Similarly, using Raman spectroscopy on $HgBa_2CuO_{4+x}$, Le Tacon *et al.* made the same identification of distinct energy scales found in antinodal and nodal spectra[31].

It is reasonable to ask why, after so many years of high resolution STM spectroscopy on a wide variety of high-$T_C$ materials, the superconducting gap may only have been revealed now. One important reason is our newly constructed STM, which is the first capable of making temperature dependent measurements while maintaining a constant position on the sample. This ability is necessary for our normalization technique. Another reason lies in our choice of sample, Bi-2201, where the energy scales of the small and large gaps are well enough separated that they may both be clearly resolved. This suggests an important future direction, namely replication of this study at other dopings and on other materials. In particular, it is important to measure doping dependence in the underdoped regime, where STM and ARPES measured gap widths typically diverge from other measurements[20], and verify that these two gaps follow two different energy scales (superconducting, $T_C$, and pseudogap, $T^*$), as indicated by new Raman[31] and ARPES[29] measurements. Of course, it remains an open question whether two distinct gaps exist in other materials at all, but in retrospect, the subgap kink which has been ubiquitously reported in STM measurements strongly suggests that they do[15,26,27].

One might also wonder if interpretations of the wide variety of previous STM results which viewed "the gap" as the superconducting gap are completely incorrect. Although they should be revisited in light of our findings, it seems likely that many will stand, given the closeness of pseudogap and superconducting gap energy scales in the most commonly studied material, Bi-2212 near optimal doping. Further experiments will be needed to determine if this similarity of energy scales means that superconducting gap properties dominate behavior of "the gap" and hence are reasonably associated with it, as has previously been done. On the other hand, results focusing on nanoscale inhomogeneity should surely be reevaluated, as should results on underdoped materials. This may further resolve apparent conflicts between STM and other measurements, leading to a more unified experimental picture of high temperature superconductivity in general.

Our present findings also represent a unifying step in the two gap picture of HTS in particular, where a large, spatially inhomogeneous gap (the pseudogap) opens near the antinode above $T_C$ and then coexists with a smaller, spatially homogeneous gap (the superconducting gap) that opens near the node at $T_C$. The opening of this sharp gap at $T_C$ may prove difficult to explain by "one gap" theories[5-11], in which the pseudogap and superconducting states differ mostly in the onset of phase coherence, and seems more in line with theories in which the pseudogap is due to some other, possibly competing, phase, in which the magnitude of the pseudogap and superconducting gaps are not directly related[10,27,32-40].

**Acknowledgements**

We thank J.C. Davis, J.E. Hoffman, P.A. Lee, Y.S. Lee, K. McElroy, T. Senthil, Y. Wang and X.-G. Wen for comments. This research was supported in part by the MRSEC Program of the National Science Foundation and by an NSF CAREER award.

**Author Contributions:**

MCB, WDW and KC shared equal responsibility for all aspects of this project from instrument construction through data collection and analysis. MY conceived the temperature normalization scheme and performed much of the data analysis. TK grew the samples and helped refine the STM. TT and HI contributed to sample growth. EWH advised.

Correspondence and requests for materials should be addressed to E.W. Hudson, ehudson@mit.edu